\documentclass[preprint,aps,nofootinbib,tightenlines]{revtex4}

\usepackage{psfig}

\begin{document}

\begin{titlepage}

\begin{flushright}
\vbox{
\begin{tabular}{l}
TPI-MINN-02/48\\
UMN-TH-2122/02\\
hep-ph/0212231
\end{tabular}
}
\end{flushright}

\vspace{1cm}

\title{%
Perturbative and nonperturbative renormalization\\ 
of anomalous quark triangles
}

\author{Arkady Vainshtein}
\affiliation{Theoretical Physics Institute, University of Minnesota,
116 Church St.\ SE, Minneapolis, MN 55455 \vspace{1.5cm}}

\begin{abstract}

\vspace{0.5cm}

Anomalous quark triangles with one axial and two vector currents
are studied in special kinematics when one of the vector currents
carries a soft momentum. According to the Adler-Bardeen theorem the
anomalous  longitudinal  part of the triangle is not
renormalized in the chiral limit.
We derive a new nonrenormalization theorem for the transversal part
of the triangle. This nonrenormalization, in difference with the
longitudinal part, holds on only perturbatively. At the nonperturbative
level we use the operator product expansion and the pion dominance in
the longitudinal part to determine the magnetic susceptibility
of the quark condensate, $\chi=-N_c/(4\pi^2\,F_\pi^2)\,$. 
\end{abstract}
 
\maketitle
\thispagestyle{empty}
\end{titlepage}

\newpage

\section{Introduction}

Study of fermion triangle diagrams with one axial and two vector
currents represents a remarkable story. The Adler-Bell-Jackiw anomaly
in the divergence of axial current, the Adler-Bardeen
nonrenormalization theorem, the Wess-Zumino effective action,
calculation of the $\pi^0 \to\gamma \gamma $ amplitude, the 't Hooft
consistency condition, and the solution of the U(1) problem give an
incomplete list of acts where these triangles played a major role.

The famous  Adler-Bardeen theorem \cite{Adler:er} proves
nonrenormalization for the longitudinal part of
triangles associated with the divergence of the axial current. There
is no general statement about the transversal part of the
triangle. This part, even its existence, depends on the choice of
external momenta. In this note we argue that in special kinematics when 
one of the vector currents carries a soft momentum the  transversal part
is unambiguously fixed by the longitudinal one in the chiral limit of
perturbation theory. Such relation immediately proves an absence of
perturbative corrections to the transversal part of fermion triangles
in the kinematics considered.

A particular physical situation where the triangle diagrams enter in
 special kinematics with one soft momentum occurs for the two-loop
electroweak corrections to the muon anomalous magnetic moment. 
In Ref.\,\cite{CMV} --- the study which  stimulated the present work --- 
the nonrenormalization theorem for the transversal part is used to
show an absence of gluon corrections to light quark loops.

The difference between longitudinal and transversal parts shows up at
a nonperturbative level. The Operator Product Expansion (OPE)
demonstrates this explicitly: in the chiral limit only the transversal
part in the OPE contains nonleading operators \cite{CMV}.
Nonrenormalization of the longitudinal part, both perturbatively and
nonperturbatively, constitutes the 't~Hooft consistency condition
\cite{tHooft}, i.e.\ the exact quark-hadron duality. In QCD this
duality is realized as a correspondence between the infrared
singularity of the quark triangle and the massless pion pole in terms
of hadrons.\footnote{A pioneering effort to analyze the axial current
anomaly in terms of infrared singularity was made by Dolgov and
Zakharov \cite{DZ}.}
It is clear that for the transversal part this kind of duality cannot
be exact in QCD: there is no massless particle contributing to the
transversal part.  Thus, the transversal part of the triangle with a
soft momentum in one of the vector currents provides us with an
interesting object: no perturbative corrections but nonperturbatively
it is modified.

An example of a nonperturbative quantity related to fermion triangles
is the quark condensate magnetic susceptibility. It was introduced in
Ref.\,\cite{Ioffe:1984ju} via the matrix element of $\bar q\,
\sigma_{\alpha\beta}q$ between the vacuum and soft photon states.
Using the OPE analysis of the triangle amplitude carried on in
Ref.\,\cite{CMV} together with an additional assumption about the pion
dominance in the longitudinal part we will derive a new relation for
the magnetic susceptibility. This relation, similar to the
Gell-Mann-Oakes-Renner relation between the pion and quark masses, is
in agreement with the QCD sum rule fit \cite{Ioffe:1984ju}.

\section{Hadronic corrections to quark triangles}
\label{sec:had}

We follow Ref.\,\cite{CMV} in notations and definitions. Let us 
 start with a definition of  vector, $j_\mu\,$, and axial,  $j^5_\nu$,
currents,
\begin{eqnarray}
j_\mu=\bar q \,V \gamma_\mu q, \qquad 
j^5_\nu = \bar q \,A \gamma _\nu \gamma _5 q\,,
\end{eqnarray}
where the quark field $q^i_f$ has color ($i$) and flavor ($f$) indices
and the matrices $V$ and $A$ are diagonal matrices of vector and axial
couplings acting on flavor indexes.  To avoid dealing with the U(1)
anomaly in respect to gluon interactions we assume that Tr$\,A=0\,$.
In the case of electroweak corrections one can view the vector current
as an electromagnetic one with $V$ being the matrix of electric
charges and $j^5_\nu$ as the axial part of the $Z$ boson current with
matrix $A$ given by the weak isospin projection.

\begin{figure}[ht]
\hspace*{0mm}
\begin{minipage}{16.cm}
\begin{tabular}{c@{\hspace{16mm}}c}
\psfig{figure=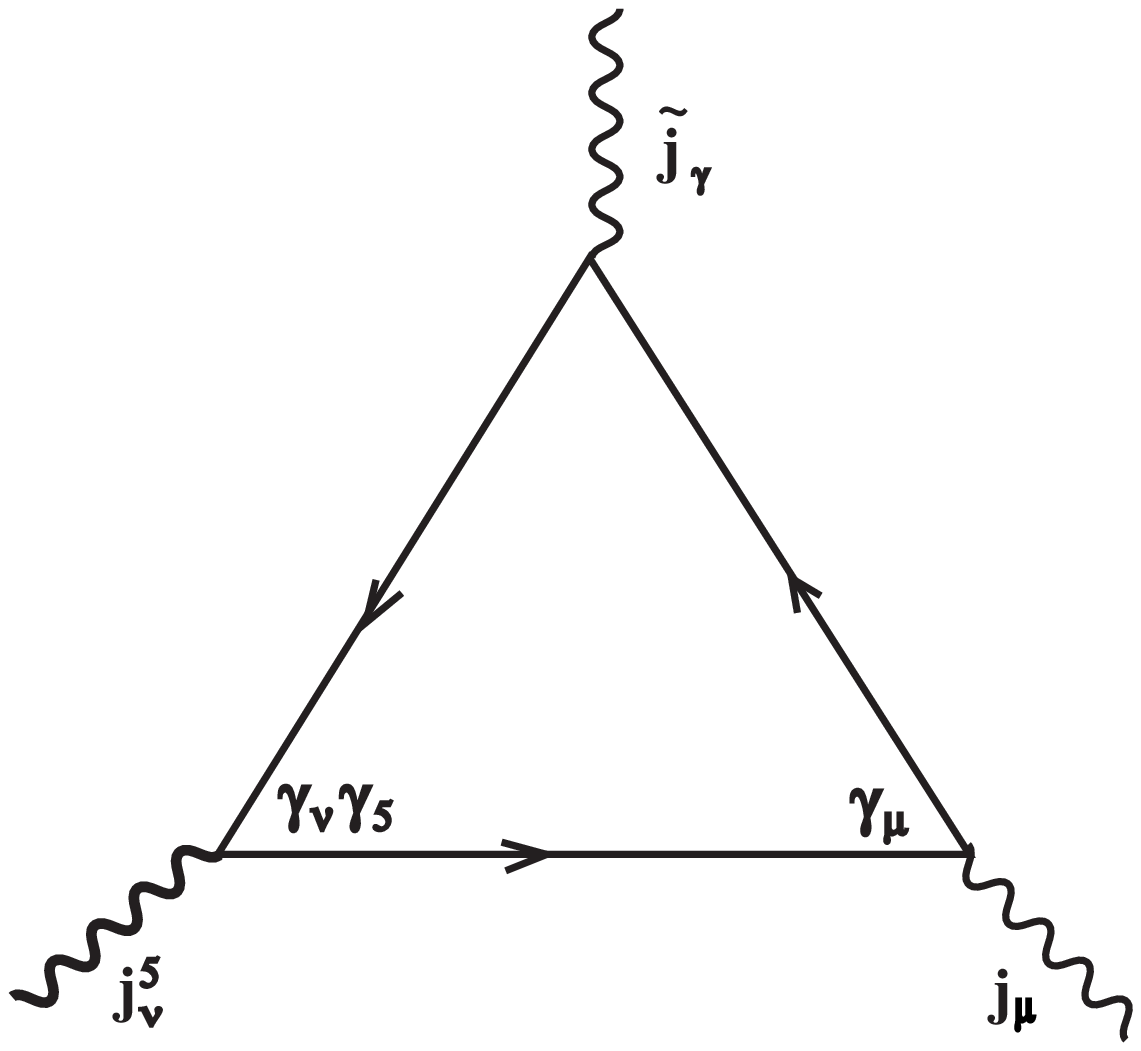,width=37mm}
&
\psfig{figure=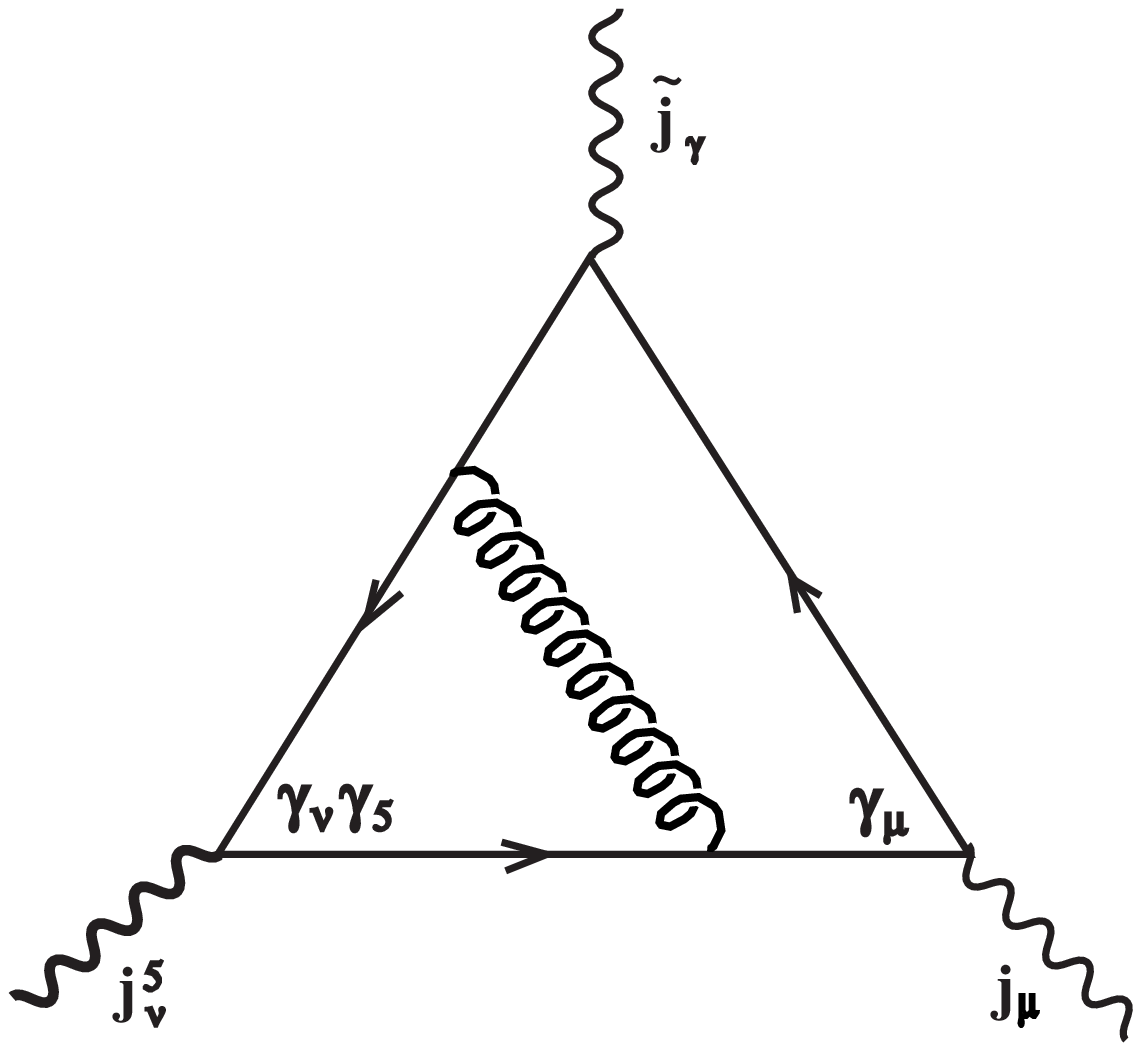,width=37mm}
\\
$(a)$  & $(b)$\\[1mm]
\end{tabular}
\end{minipage}
\caption{\sf \small Quark triangle, diagram $(a)$, and a gluon 
  correction to it, diagram $(b)\,$.} 
\label{fig:triangle}
\end{figure}

The amplitude for the triangle diagram in Fig.\,\ref{fig:triangle}
 involving the axial current $j^5_\nu$
and two vector currents 
$j_\mu$ and $\tilde j_\gamma=\bar q \,\widetilde V \gamma_\mu q$ 
(for generality we use a different matrix $\widetilde V$ for 
the soft momentum current) can be written as
\begin{equation}
T_{\mu \gamma \nu } =-\int {\rm d}^4 x {\rm d}^4 y\, 
{\rm e}^{iqx-iky}\,\langle 0|\,T\{j_\mu(x)\, \tilde j_\gamma(y)\,  
j^5_\nu (0)\}|0\rangle\,.
\end{equation}
We can view the current $\tilde j_\gamma$ as a source of a soft photon
with the momentum $k\,$. Introducing a polarization vector of a soft
photon $e^\gamma(k)$ we come to the amplitude
 $T_{\mu \nu} $
\begin{equation}
T_{\mu \nu }=T_{\mu \gamma \nu } e^\gamma(k) =i\int {\rm d}^4 x\, 
{\rm e}^{iqx}\,\langle 0|\,T\{j_\mu
(x)\, j^5_\nu (0)\}|\,\gamma (k)\rangle\,,
\end{equation}
which can be viewed as a mixing between the
 axial and vector currents in the external electromagnetic field.
 
It is clear that the expansion of $T_{\mu \nu }$ in the small
momentum $k$ starts with linear terms and we neglect quadratic and
higher powers of $k$. There are only two Lorentz structures for
$T_{\mu \nu }$ which are linear in $k$ and consistent with the
conservation of electromagnetic current,
\begin{eqnarray}
\label{invfun}
&&T_{\mu \nu }=-\frac{i}{4\pi^2}\left[w_T(q^2) 
\left(-q^2 \tilde f_{\mu \nu
}+q_\mu q^\sigma
\tilde f_{\sigma  \nu } - q_\nu q^\sigma \tilde f_{\sigma  \mu }\right)
+w_L(q^2)\, q_\nu q^\sigma \tilde f_{\sigma  \mu }\right],\\[1mm]
&&\tilde f_{\mu \nu }=\frac{1}{2}\, \epsilon _{\mu \nu 
\gamma
\delta }  f^{\gamma \delta }\,,\qquad  f_{\mu \nu }=k_\mu e_\nu -k_\nu
e_\mu
\,.\nonumber
\end{eqnarray}
Both structures are transversal with respect to vector current,
$q^\mu T_{\mu\nu}=0$. As for the axial current, the first structure is
transversal with respect to $q^\nu$ while  the second is
longitudinal.

The one-loop result for the invariant functions $w_{T}$ and $w_L$ can
be taken from the classic papers by Bell and Jackiw~\cite{BJ},
Adler~\cite{Adler:1969gk} and Rosenberg~\cite{Rosenberg:1963pp} (it
simplifies considerably in the limit of the small photon momentum
\cite{KKSS}),
\begin{equation}
\label{wlt}
w^{\rm 1-loop}_L=2\,w^{\rm 1-loop}_T=2N_c\,{\rm Tr} \,A\,V\,
\widetilde V \int_0^1
\frac{{\rm d}\alpha\, \alpha (1-\alpha )}{\alpha (1-\alpha )Q^2+m^2}
\,,
\end{equation}
where $Q^2=q^2\,$, the factor $N_c$ accounts for the color of quarks
and $m$ is the diagonal quark mass matrix, 
$m=\mbox{diag}\{m_{q_1}, m_{q_2}, \ldots\}$.
In the chiral limit, $m=0$, the invariant functions $w_{T,L}$ are
\begin{equation}
\label{msquare}
w^{\rm 1-loop}_L[m=0]=2\,w^{\rm 1-loop}_T[m=0]
=\frac{2N_c\,{\rm Tr} \,(A\,V\,\widetilde V)}{Q^2}\,.
\end{equation}
Nonvanishing  in the chiral limit, $m=0\,$, the longitudinal part
$q^\nu T_{\mu\nu}$ represents the axial anomaly \cite{BJ,Adler:1969gk},
\begin{equation}
\label{anomaly}
q^\nu T_{\mu\nu}=\frac{i}{4\pi^2} \,Q^2 w_L\,q^\sigma \tilde
f_{\sigma\mu} 
=\frac{i}{2\pi^2} \,N_c\,{\rm Tr} \,(A\,V\,\widetilde V)\,
q^\sigma \tilde f_{\sigma\mu}\,,
\end{equation}
and its nonrenormalization implies that the one-loop result
(\ref{msquare}) for $w_L$ stays intact when interaction with gluons
is switched on.

\subsection{Nonrenormalization theorem for the transversal part of the
  triangle}

We claim that the relation
\begin{equation} 
w_L[m=0]=2\,w_T[m=0]
\label{wtwl}
\end{equation}
which holds at the one-loop level, see Eq.\,(\ref{msquare}), gets
no perturbative corrections from gluon exchanges. This
follows from the following line of argumentation. 

In the chosen kinematics the fermion triangle with $m=0$ possesses a
special feature: namely, a symmetry under permutation of indexes of
axial and vector currents, $\mu \leftrightarrow \nu\,$. Indeed, in the
triangle diagrams $(a)$ and $(b)$ in Fig.\,\ref{fig:triangle} one can
move $\gamma_5$ from the axial vertex $\gamma_\nu \gamma_5$ to the
vector vertex $\gamma_\mu\,$. In the chiral limit it moves via even
number of gamma matrices in any order of perturbation theory. Together
with the change of the momentum $q \to -q$ (which does not affect
$T_{\mu\nu}$) it shows the symmetry of the amplitude $T_{\mu\nu}\,$.

At first glance the symmetry under the $\mu \leftrightarrow \nu$
permutation seems to be in contradiction with the general
decomposition (\ref{invfun}): the transversal part of $T_{\mu\nu}$ is
antisymmetric, the longitudinal part has no symmetry, and there is no
way to choose $w_T$ and $w_L$ which makes the $T_{\mu\nu}$ symmetric.
Note, however, that the term $q^2 \tilde f_{\mu\nu}$ in the
transversal structure in Eq.\,(\ref{invfun}) actually produces a term
in $T_{\mu\nu}$ which does not depend on $q$. It is because
$w_T\propto 1/q^2$. The $\mu \leftrightarrow \nu$ symmetry holds for a
singular in $q$ part of $T_{\mu\nu}$ when the condition (\ref{wtwl})
relating $w_T$ to $w_L$ is fulfilled. The constant in $q$ part is then
fixed by the conservation of the vector current, $q^\mu T_{\mu\nu}=0\,$. 
An independence on $q$ for the antisymmetric part
provides, in fact, an alternative proof of the Adler-Bardeen theorem.
Indeed, gluon corrections would lead to logarithmic dependence on $q$
instead of the constant.

Another way to be automatically consistent with
the vector current conservation is to use the Pauli-Vilars
regulators. Technically it reduces to subtraction from the triangles
with massless quarks  similar triangles with the heavy regulator
fermions propagating on the loops. The regulator triangles produce
terms which are polynomial in momenta, in our case terms linear in $k$
and independent on $q\,$. Moreover, it is simple to see that these
terms are antisymmetric under the $\mu \leftrightarrow \nu$
permutation. Indeed, in the propagator of the heavy regulator the
leading term contains no gamma-matrix that leads to the sign change
when $\gamma_5$ from the axial vertex $\gamma_\nu
\gamma_5$ is moved to the vector vertex $\gamma_\mu\,$.

Thus, we see that the crossing symmetry of the singular part in the
triangle amplitude $T_{\mu\nu}$ leads to the relation (\ref{wtwl}) in
perturbation theory. Nonrenormalization of $w_L$ implies the same
for $w_T\,$.

\subsection{Nonperturbative effects and OPE}
\label{sec:OPE}

To study a nonperturbative effect in the triangle amplitude
$T_{\mu\nu}$ one can use the OPE methods.  This section is a brief
review of the OPE analysis made in Ref.\,\cite{CMV}. The analysis
shows a nonpertubative difference between the longitudinal and
transversal parts, we will use the results in the next section.

The OPE  for the T-product of electromagnetic and axial currents
at large Euclidean $q^2$ has the form
\begin{equation}
\label{ope}
\hat T_{\mu \nu }\equiv i\int {\rm d}^4 x\, {\rm e}^{iqx}\,
T\{j_\mu (x)\, j^5_\nu (0)\}=
\sum_i c^i_{\mu \nu \gamma_1\ldots\gamma _i}(q)\,
{\cal O}_i^{\gamma_1\ldots\gamma _i}\,,
\end{equation}
where the local operators ${\cal O}_i^{\gamma_1\ldots\gamma _i}$ are
constructed from the light fields and supplied by a normalization
point $\mu$ separating short distances (accounted in the coefficients
$c^i$) and large distances (in matrix elements of ${\cal O}_i$).  The
field can be viewed as light if its mass is less than $\mu$. In the
problem under consideration besides quark and gluon fields this
includes also  the soft electromagnetic field $A_\mu\,$.
 The field $A_\mu$ could enter local operators in a form of the gauge
invariant field strength $F_{\mu \nu }=\partial _\mu A_\nu -\partial
_\nu A_\mu $. 

The amplitude $T_{\mu \nu }$ is given by the matrix element of 
the operator $\hat T_{\mu \nu }$ between the photon and vacuum states,
\begin{equation}
\label{opemat}
T_{\mu \nu }=\langle 0|\,\hat T_{\mu\nu }\,|\gamma (k)\rangle=
\sum_i c^i_{\mu \nu \alpha_1\ldots\alpha_i}(q)\,
\langle 0|\,{\cal O}_i^{\alpha_1\ldots\alpha_i}\,|\gamma (k)\rangle\,.
\end{equation}
In our approximation, when the matrix
elements are linear in $f_{\alpha \beta }\!=\!k_\alpha e_\beta\! 
-\!k_\beta e_\alpha\,$, they are nonvanishing only for operators with 
a pair of antisymmetric indexes,
\begin{equation}
\langle 0|\,{\cal O}_i^{\alpha\beta}\,|\gamma 
(k)\rangle=-\frac{i}{4\pi^2}\,\kappa
_i\,\tilde f^{\alpha\beta}\,,
\end{equation}
where constants $\kappa _i$ depend on the normalization point $\mu\,$.
With only contributing operators the OPE takes the form
\begin{equation}
\hat T_{\mu \nu }\!=\!\sum_i\left\{
c^i_T(q^2) \!\left(-q^2 {\cal O}^i_{\mu \nu}\!+\!q_\mu q^\sigma
{\cal O}^i_{\sigma  \nu }\! -\! q_\nu q^\sigma {\cal O}^i_{\sigma  \mu 
}\right)
\!+\!c^i_L(q^2)\, q_\nu q^\sigma {\cal O}^i_{\sigma  \mu}\right\},
\end{equation}
and the invariant functions $w_{T,L}$ can be presented as
\begin{equation}
w_{T,L}(q^2)=\sum c^i_{T,L}(q^2)\, \kappa_i\,.
\label{kappa}
\end{equation}

The leading (by a minimal dimension) is the $d=2$ operator 
\begin{equation}
{\cal O}_F^{\alpha\beta}=\frac{1}{4\pi^2}\,\tilde F^{\alpha\beta}
=\frac{1}{4\pi^2}\,
\epsilon^{\alpha\beta \rho \delta }\partial
_\rho A_\delta\,,
\label{opF}
\end{equation}
where $\tilde F^{\alpha\beta}$
is the dual of the electromagnetic field strength. The numerical
factor in (\ref{opF}) is such that $\kappa_F=1\,$.
The OPE coefficients for ${\cal O}_F^{\alpha\beta}$ follow 
from one-loop expressions (\ref{wlt})
for $w_{L,T}\,$,
\begin{equation}
\label{cF}
c^F_{L}[\mbox{1-loop}]=2c^F_{T}[\mbox{1-loop}]=
\frac{2\,N_c}{Q^2}\,{\rm Tr}\,A\,V\widetilde
V\!\left[1
+{\cal O}\left(\frac{m^2}{Q^2}\right)\right],
\end{equation}
where we imply that  $m \ll \mu \ll Q\,$, see \cite{CMV} for a more
detailed discussion.

The next, by dimension, are $d=3$ operators 
\begin{equation}
{\cal O}^{\alpha\beta}_{f}=
-i\,\bar q_f\,\sigma^{\alpha\beta}\gamma _5\, q^f 
\equiv \frac{1}{2}\,\epsilon^{\alpha\beta\gamma\delta}
\bar q_f\,\sigma_{\gamma\delta}\, q^f\,,
\label{fop}
\end{equation}
where the index denotes the quark flavor. The OPE coefficients follow
from  tree diagrams of the Compton scattering type,
\begin{equation}
c^f_L=2c^f_T=\frac{4\,A_f V_f \,m_f}{Q^4}\,.
\end{equation}
Proportionality to $m_f$ is in correspondence with chirality
arguments.
 Taking matrix element of ${\cal O}^{\alpha\beta}_{f}$ between 
the soft photon and vacuum
states we produce 
the following terms in the invariant functions $w_{T,L}(q^2)$:
\begin{eqnarray}
\label{d3}
\Delta^{(d=3)} w_L=2\,\Delta^{(d=3)}w_T=\frac{4}{Q^4}\sum_f A_f V_f m_f
\kappa_f \,.
\end{eqnarray}

In perturbation theory the matrix element $\kappa_f$ of the
chirality-flip operator $O_f$ is proportional to $m_f$.
Nonperturbatively, however, $\kappa_f$ does not vanish at
$m_f\!=\!0\,$.  Due to spontaneous breaking of the chiral symmetry in
QCD the matrix elements of quark operators (\ref{fop}) are instead
proportional to the quark condensate 
$\langle \bar q q\rangle_{_0}=-(240\,{\rm MeV})^3\,$.  
It leads to the representation of $\kappa_f$ in the form
\begin{eqnarray}
\label{magmom}
\kappa_f=-4\pi^2 \widetilde V_f\,\langle \bar q
  q\rangle_{_0}\,\chi\,.
\end{eqnarray}
This representation was introduced by Ioffe and Smilga
\cite{Ioffe:1984ju} in their 
analysis of nucleon magnetic moments and $\Delta \to N \gamma$
radiative transitions with QCD sum
rules.  From a sum rule fit they determined the value of the parameter
$\chi$ dubbed as the quark condensate magnetic susceptibility,
\begin{eqnarray}
\label{magmom1}
\chi=-\frac{1}{(350 \pm 50~\mbox{MeV})^2}\,.
\end{eqnarray}

We will consider an analytical calculation and comparison with other
approaches for the susceptibility $\chi$ in the next section. Here we
notice that the effect of $d=3$ operators ${\cal O}_f$ vanishes in the
chiral limit, although as the first rather than second power of $m\,$.
What we are looking for, first of all, are terms in the OPE which
differentiate longitudinal and transversal parts in this limit.

Vanishing at the chiral limit persists also for the $d=4$ and $d=5$
operators.  All operators of dimension 4 are reducible to the $d=3$
operators due to the following relation,
\begin{equation}
 \bar q_f \,(D_\mu \gamma _\nu -
D_\nu \gamma _\mu)\gamma _5\, q^f=-m_f \,\bar q_f 
\sigma_{\mu \nu}\gamma_5\,q^f\,.
\end{equation}
The $d=5$ operators of the type 
$\bar q_f q^f \tilde F^{\alpha\beta}$ and $\bar q_f\gamma _5 q^f \tilde
F^{\alpha\beta}$ enter OPE with  factors $m_f$  as in the $d=3$ case.

The distinction between longitudinal and transversal parts shows up at
the $d=6$ level of four-fermion operators as it was firstly
demonstrated in Ref.\,\cite{KPPR}.  Referring to Ref.\,\cite{CMV} for
more detailed discussion note here that these four-fermion operators change
the transversal, but not longitudinal, part. Arising due to these
operators terms $1/Q^6$ in $w_T$ reflect nonvanishing masses of meson
resonances contributing to the transversal part. This was used in
\cite{CMV} for construction of a resonance model for $w_T\,$
consistent with the OPE constraints.

\section{Pion dominance and  magnetic 
susceptibility of quark condensate}

In this section we limit ourselves by  the axial
current  of the light $u$ and $d$ quarks, 
$j_\nu^5=\bar u\gamma_\nu \gamma_5 u - \bar d\gamma_\nu \gamma_5 d$, 
i.e.\ $A=\mbox{diag}\{A_u,A_d\}=\mbox{diag}\{1,-1\}$.

A specific feature of the longitudinal part of $T_{\mu\nu}$ in the
chiral limit is  that it is given by the leading $d=2$ operator
$\tilde F_{\alpha\beta}$ in the
whole range of $Q$, from the ultraviolet to infrared,
\begin{equation}
\label{wlexa}
w_L[m_{u,d}=0]=\frac{2N_c\,{\rm Tr} \,(A\,V\,\widetilde V)}{Q^2}\,.
\end{equation}
At large $Q$ it is fixed by the OPE, the pole singularity at small $Q$ is
due to massless pion with the residue that matches the OPE. How is
$w_L$ changed at nonvanishing but small $m_{u,d}\,$?

A nonvanishing $m_f$ implies a nonvanishing pion mass so the pole
in $w_L$ should be shifted from zero,
\begin{equation}
\label{wlmne0}
w_L[m_{u,d}\neq 0]=\frac{2N_c\,{\rm Tr} \,(A\,V\,\widetilde V)}{Q^2
+m_\pi^2}\,.
\end{equation}
This expression extends the pion pole dominance, which is exact at
$m_{u,d}=\,0$, to the case of  small but nonvanishing  $m_{u,d}\,$. Such
extension is certainly valid for $Q$ which is much smaller
than the characteristic hadronic scale, say the $\rho$ meson mass
$m_\rho\,$. Moreover,  at large $Q$   the leading
$1/Q^2$ term  in Eq.\,(\ref{wlmne0}) matches what follows from the
OPE. It is not enough, strictly speaking, to justify the pion
dominance for the next, $m_\pi^2/Q^4$, term in expansion at large $Q\,$.

Assuming that the pion pole dominance for the $1/Q^4$ term holds --- we will
return to this later --- we can compare it with what follows from the OPE.
From Eq.\,(\ref{wlmne0})  
the coefficient of $1/Q^4$ is
\begin{equation}
\label{expan }
-2m_\pi^2 \,N_c\,(V_u\,\widetilde V_u- V_d\,\widetilde V_d)\,,
\end{equation}
while the OPE relation (\ref{d3}) gives for this coefficient
\begin{equation}
\label{d3coef }
4\,(V_um_u\kappa_u-V_dm_d \kappa_d)=
2(m_u+m_d)(V_u\kappa_u-V_d\kappa_d)+
2(m_u-m_d)(V_u\kappa_u+V_d\kappa_d)\,.
\end{equation}
The $m_u-m_d$ part is not relevant to comparison: it corresponds to 
the mixing of  pion with massive isoscalar states, a reflection of 
the U(1) anomaly in the linear in $m_{u,d}$ terms.
Keeping in mind that $V_{u,d}$ can be chosen arbitrarily we get 
\begin{eqnarray}
  \label{magsuc}
  (m_u+m_d)\,\kappa_f=- m_\pi^2\,N_c\,\widetilde V_f\,,
\end{eqnarray}
where $f=u,d$ and $\widetilde V_f$ is the electric charge of the $f$ quark.
This looks analogous to the Gell-Mann-Oakes-Renner (GMOR) relation for
the pion  mass \cite{GMOR},
\begin{equation}
\label{gmor}
 (m_u+m_d)\langle \bar q q\rangle_{_0}=-F_\pi^2 \,m_\pi^2\,.
\end{equation}
The GMOR relation allows us to rewrite (\ref{magsuc}) as a relation for 
the magnetic susceptibility $\chi$ defined in Eq.\,(\ref{magmom}),
\begin{eqnarray}
  \label{magsuc1}
\chi=-\frac{N_c}{4\pi^2 \,F_\pi^2} =-\frac{1}{(335~\mbox{MeV})^2} \,.
\end{eqnarray}

The $N_c$ dependence of the result for $\chi$ is consistent with large
$N_c$ analysis. The numerical value of $\chi$ is in very good
agreement with the QCD sum rule fit \cite{Ioffe:1984ju} given in
Eq.\,(\ref{magmom1}). What remains questionable is the pion
dominance, which we will discuss in a little bit more detail.

To this end it is instructive to compare the construction above with
the OPE derivation of the GMOR relation (\ref{gmor}) made in
Ref.\,\cite{SVZ}. The object of consideration in this case was the
polarization operator $\Pi_{\mu\nu}$ of the axial current $j_\mu^5$.
In its longitudinal part the $d=3$ operators give
\begin{equation}
\label{piax }
\Delta \Pi_{\mu\nu}^{(d=3)}
=2(m_u+m_d)\langle \bar q q\rangle_{_0}\frac{q_\mu q_\nu}{q^4}
+\mbox{transversal terms}\,.
\end{equation}
Comparing this with the $m_\pi^2/q^4$ term coming from the expansion
in the pion pole one gets the GMOR relation (\ref{gmor}).  It is
crucial that only the pion state contributes to the linear in $m_{u,d}$
(or in $m_\pi^2)$ terms in the longitudinal part of $\Pi_{\mu\nu}$,
all the higher states give quadratic in quark masses contributions.
Indeed, the coupling of those states to the axial current is linear in
quark masses and it is the square of this coupling which enters
$\Pi_{\mu\nu}\,$.  In the case of the longitudinal part of $T_{\mu\nu}$
the coupling of higher states to the axial current enters only once,
so the higher states do contribute in the linear in quark masses
order. Thus, the pion dominance is not parametrical for $q^\nu
T_{\mu\nu}\,$.

A clear signal of presence of higher states follows from a nonvanishing 
anomalous dimension of the operator (\ref{fop}). It means that the
operator (\ref{fop}) (in contrast with the operator 
$(m_u+m_d)\, \bar q q$ entering the GMOR relation (\ref{gmor})) depends
on the normalization point $\mu$, and its OPE coefficient $c_L^f (Q)$
besides power dependence on $Q$ contains also the factor
$[\alpha(Q)/\alpha(\mu)]^{16/9}$, i.e.\ power of 
$\log(Q/\Lambda_{\rm QCD})\,$.
This logarithmic dependence is apparently related to the higher
states contribution. To justify the pion dominance we have to assume 
matching of
the $1/Q^4$ terms from the OPE and the pion pole below the higher
states. It implies a low normalization point, probably $\mu\sim
0.5~\mbox{GeV}$. 

The result (\ref{magsuc1}) can be compared with a different approach
to calculation of $\chi$ based on matching of the vector meson dominance
with the OPE for the product of the electromagnetic current 
$\tilde j_\gamma$ and operator (\ref{fop}). This approach was
suggested first in Ref.~\cite{BY} and in its simplest form gives
$\chi=-2/m_\rho^2=-1/(544~{\rm MeV})^2$ what is about 2.6 times
smaller by magnitude than (\ref{magsuc1}). The consideration was then 
improved in \cite{BK,BKY} by use of the QCD sum rules, see 
also \cite{BBK} for a recent review and update.  The largest by magnitude 
value $\chi[\mu\!=\!0.5~\mbox{GeV}]=-1/(420~{\rm MeV})^2$ obtained in
\cite{BK} is still 1.5 times smaller than (\ref{magsuc1}). A
phenomenology of processes sensitive to the susceptibility $\chi\,$, 
see \cite{BBK}, will possibly help to fix its value.

\newpage
\section{Conclusions}

The quark triangles in special kinematics with one soft photon are
similar to polarization operators: in this case it is a nondiagonal
mixing of the axial and vector currents in the background of a soft vector
field. In this kinematics we find that perturbative gluon corrections
are absent in the chiral limit not only for the longitudinal part but
for the transversal part as well. At the nonperturbative level the
transversal part is corrected in contrast with the longitudinal one.
In this respect the hadronic shift in the transversal part represents
an object similar to the quark condensate: it appears only at
nonperturbative level.

We also derive a new expression for the quark condensate magnetic
susceptibility using the OPE and pion dominance in the longitudinal
part of T-product of axial and two vector currents. This expression is
similar to the Gell-Mann-Oakes-Renner relation between pion and quark
masses which also can be derived by the OPE method.  The crucial
difference is, however, that the pion dominance is parametrically
valid for the GMOR relation but not for the magnetic susceptibility.
Although theoretical accuracy of the new relation is not clear it
would be interesting to follow further its phenomenological
consequences.

\begin{acknowledgments} 
I am grateful to S.~L.~Adler, V.~M.~Braun, A.~Czarnecki, G.~Gabadadze,
E.~D'Hoker, W.~J.~Marciano, M.~Shifman and M.~Voloshin for helpful
discussions and am thankful for the hospitality of the Aspen Center
for Physics where a part of this work was done.

The work was supported in part by  DOE grant DE-FG02-94ER408.
\end{acknowledgments}

\end{document}